\documentclass[10pt,aps,prd,nofootinbib,superscriptaddress,twocolumn,preprintnumbers,balancelastpage]{revtex4}
\usepackage{amssymb,amsmath,latexsym,graphics, graphicx,epsfig,multirow,comment,hyperref,appendix,feyn,slashed} 

\usepackage{color}

\newcommand{\overbar}[1]{\mkern 2mu\overline{\mkern-2mu#1\mkern-2mu}\mkern 2mu}

\newcommand {\be} {\begin {equation}}
\newcommand {\ee} {\end {equation}}

\newcommand {\bes} {\begin {equation*}}
\newcommand {\ees} {\end {equation*}}

\newcommand{\es}[2] {\begin{equation} \label{#1} \begin{split} #2 \end{split} \end{equation}}

\newcommand{\beq}{\begin{equation}}
\newcommand{\eeq}{\end{equation}}

\newcommand {\kms} {\,\,\text{km}/\text{s}}

\begin{document}

\title{
Measuring Anisotropies in the Cosmic Neutrino Background 
 }
\author{Mariangela Lisanti}
\affiliation{Department of Physics, Princeton University, Princeton, NJ 08544}
\author{Benjamin R. Safdi}
\affiliation{Department of Physics, Princeton University, Princeton, NJ 08544}
\author{Christopher~G. Tully}
\affiliation{Department of Physics, Princeton University, Princeton, NJ 08544}

\date{\today}

\begin{abstract}
Neutrino capture on tritium has emerged as a promising method for detecting the cosmic neutrino background (C$\nu$B).  We show that relic neutrinos are captured most readily when their spin vectors are anti-aligned with the polarization axis of the tritium nuclei and when they approach along the direction of polarization.
As a result, C$\nu$B observatories may measure anisotropies in the cosmic neutrino velocity and spin distributions by polarizing the tritium targets. 
A small dipole anisotropy in the C$\nu$B is expected due to the peculiar velocity of the lab frame with respect to the cosmic frame and due to late-time gravitational effects.  The PTOLEMY experiment, a tritium observatory currently under construction, should observe a nearly isotropic background.  This would serve as a strong test of the cosmological origin of a potential signal.  The polarized-target measurements may also constrain non-standard neutrino interactions that would induce larger anisotropies and help discriminate between Majorana versus Dirac neutrinos.
\end{abstract}
\maketitle

The cosmic neutrino background (C$\nu$B) formed when neutrinos decoupled from the thermal Universe nearly one second after the Big Bang~\cite{Weinberg:2008zzc}.
Today, these relic neutrinos are predicted to have a temperature of \mbox{$T_\nu \approx 1.95$ K}~\cite{Dicke:1965zz}.  Because they are at least partially non-relativistic, their distribution should be gravitationally perturbed as they free-stream towards us.  As a result, a successful detection of the C$\nu$B and its anisotropies would be an astounding demonstration of early-Universe physics, while also probing late-time structure.    
 
Neutrino capture on beta decaying nuclei (NCB) is a promising path forward towards ultra-low-energy neutrino detection~\cite{Weinberg:1962zza}.  NCB has no energy threshold on the incoming neutrino, making it ideal for cosmic neutrino detection.  For non-relativistic neutrinos, the neutrino-capture electrons are separated from the beta-decay electrons by a small energy gap of order the neutrino mass.
Planck+WMAP and high-$\ell$ data constrain the sum of neutrino masses to be below $0.66$ eV (95\% C.L.), while including baryon acoustic oscillation data may tighten the bound to $0.23$ eV (95\% C.L.)~\cite{Ade:2013zuv}.  Neutrino oscillation experiments indicate that at least one mass-eigenstate has a mass greater than \mbox{$\sim$$0.05$ eV}~\cite{Beringer:1900zz}. 
 
  Cosmic neutrino direct detection is one of the outstanding problems facing modern physics, and it deserves dedicated experimental and theoretical investigation.
   PTOLEMY~\cite{Betts:2013uya} is the first experiment proposing to use NCB to detect the C$\nu$B.  Their planned target consists of $\sim$$100$ g of tritium ($^3$H) atomically bound to graphene.  PTOLEMY should observe $\sim$10 C$\nu$B capture events per year, depending on the mass hierarchy and the Dirac versus Majorana nature of the neutrinos; the rate is half as large for non-relativistic Dirac neutrinos~\cite{Betts:2013uya,Long:2014zva}.  PTOLEMY has a planned energy resolution $\sim$$0.15$ eV,  though this resolution may be further improved~\cite{Betts:2013uya}.
 
 The detection rate for relic neutrinos may be enhanced if they are clustered gravitationally.  Massive neutrinos become non-relativistic at late times, and their speeds fall below the escape speeds of galactic clusters and galaxies.  Gravitational clustering is most significant for more massive neutrinos, since these became non-relativistic at earlier times.  Simulations show that the local density of neutrinos could be enhanced over the cosmological average by an order of magnitude or more~\cite{Ringwald:2004np}.  
  
As planned, NCB experiments may observe two features of the C$\nu$B: the local density of cosmic neutrinos as well as their energies, which are expected to be equal to the neutrino mass, at least for the heaviest eigenstate, up to small thermal corrections.  The former quantity is inferred from the total rate, while the latter is obtained from the energy of the final-state electron.  
The lack of other observable quantities makes it difficult to check the cosmological origin of the signal and to learn about other features of the C$\nu$B, such as the phase-space distribution of the relic neutrinos.

In fact, the total rate may modulate throughout the year, at the $0.1$--$1$\% level, due to gravitational focusing by the Sun~\cite{Lee:2013wza,Safdi:2014rza}.  The Sun is expected to have a peculiar velocity with respect to the C$\nu$B rest frame so in the Earth's rest frame the C$\nu$B appears as a neutrino `wind.'  When the Earth is `downwind' of the Sun, the neutrinos are focused by the gravitational field of the Sun, and the local neutrino density is enhanced.  This effect is most significant for more massive neutrinos, because they have lower speeds and are thus deflected more by the Sun~\cite{Safdi:2014rza}.  An annually-modulating signal would probe the local neutrino velocity distribution.

This paper proposes a new technique that NCB experiments may use to probe anisotropies in the C$\nu$B.  If the beta-decaying target is polarized, then the capture rate is sensitive to the direction of the neutrino's spin and velocity.  In particular, the rate is maximal when the neutrino's spin is anti-parallel to that of the tritium and when its velocity is aligned with the polarization axis.  Asymmetries in the C$\nu$B lead to changes in the detection rate as the direction of the polarization axis in the sky changes.  The orientation of the detector varies throughout the day due to the rotation of the Earth.  Thus, for a fixed polarization direction on Earth, the asymmetry is manifested as a daily modulation of the rate.  
    
Standard neutrino cosmology predicts nearly isotropic spin and velocity distributions; the C$\nu$B is uniquely isotropic, compared to other neutrino and background sources.  An observatory with $\sim$100 g of tritium will not have enough target mass to observe the small C$\nu$B dipole anisotropy in the standard scenario; it should observe an isotropic flux of neutrinos.
Significant anisotropies in the rate could arise from non-standard neutrino physics.  With more target mass, the anisotropies in the C$\nu$B may be observed even in the standard cosmological scenario.  These measurements may help understand the local C$\nu$B phase-space distribution, measure the C$\nu$B temperature, and discriminate between Majorana and Dirac neutrinos.
  
This paper is organized as follows.  First, we compute the NCB cross-section on a polarized target.  Then, we discuss the implications of a polarized target for detection of the C$\nu$B.  We conclude by evaluating the feasibility of implementing this proposal at an experiment such as PTOLEMY. 

 \section{Polarized scattering amplitude}
 
 We begin by considering the neutrino capture process \mbox{$\nu_j + n \to p + e^-$}.  The generalization to the case of interest, \mbox{$\nu_j + \text{$^3$H} \to \text{$^3$He} + e^{-}$}, is straightforward.  
 The neutron and neutrino are prepared in definite spin states, while the spins of the proton and electron are not observed. 
 Here, $\nu_j$ is the neutrino in the $j^\text{th}$ mass eigenstate, which has overlap $U_{ej}$ (from the PMNS matrix) with the electron-flavor neutrino eigenstate. 
 In the Fermi theory, the NCB matrix element is 
  \es{Matrix1}{
 {\cal M} = {G_\text{F} c_1 U_{ej}^* \over \sqrt{2} }  \bar u_p \gamma_\mu (1 - g_A \gamma^5) u_n \bar u_e \gamma^\mu (1 - \gamma_5) u_{ \nu} \,,
 }
where $G_F$ is the Fermi constant, \mbox{$g_A \approx 1.27$} is the axial vector coupling, $c_1$ is the cosine of the Cabibbo angle, and $u_p$, $u_n$, $u_e$, and $u_{ \nu}$ are the free proton, neutron, electron, and neutrino wave functions, respectively.   

The amplitude for the neutrino to be captured on the neutron is then given by 
\es{M2}{
\overbar{| {\cal M}|^2} =  &{G_F^2 c_1^2 |U_{ej}|^2 \over 2}  g^{\mu \nu} g^{\rho \sigma}  A^h_{\mu \sigma} A^\ell_{\nu \rho} \,, \\
}
with 
\es{sp}{
&A^h_{\mu \sigma} = \sum_{{\bf \hat s_p} }  \bar u_p \gamma_\mu (1 - g_A \gamma^5) u_n \bar u_n \gamma_{\sigma} (1- g_A \gamma^5) u_p \,, \\
&A^\ell_{\nu \rho}(s_\nu) = \sum_{ {\bf \hat s_e}}  \bar u_e \gamma_\nu (1 -  \gamma^5) u_\nu \bar u_\nu \gamma_{\rho} (1-  \gamma^5) u_e \,.
}
The spins of the final-state electron (${\bf \hat s_e}$) and proton (${\bf \hat s_p}$) are summed over because they are not observed.

The polarized spinor products for the neutrino and the neutron are simplified using 
\es{}{
u \bar u &= {\slashed{p} + m \over 2} (1 + \gamma^5 \slashed{S} ) \\
S_\mu &= (\gamma v {\bf \hat v} \cdot {\bf \hat s}, (\gamma - 1) ( {\bf \hat v}\cdot{\bf \hat s}) {\bf \hat v} + {\bf \hat s}   ) \,,
}
with $\gamma = 1 / \sqrt{1 - v^2}$.  We evaluate the squared amplitude in the lab frame, where the neutron is at rest.  The neutron spin is taken to point in the direction ${\bf \hat s}_n$.  The incoming neutrino has velocity ${\bf v_\nu}$ and spin ${\bf  \hat s_\nu}$.  The outgoing electron has velocity ${\bf v_e}$.  The momentum of the recoiling proton can be neglected.  A straightforward evaluation then gives
\es{spinSum}{
&\overbar{ |{\cal M}|^2} = 8 G_F^2 c_1^2 |U_{ej}|^2 m_n m_p E_\nu E_e (1 + 3 g_A^2)\times \\
&\left[ 1 -  {\bf  v_\nu} \cdot {\bf \hat s_\nu}  - B\,\gamma_\nu^{-1} \, {\bf \hat s_\nu} \cdot {\bf \hat s_n} \right. \\
&\left.  +\,B\,{\bf v_\nu} \cdot {\bf \hat s_n} \left(1 - {\gamma_\nu \over \gamma_\nu + 1}{\bf  v_\nu} \cdot {\bf \hat s_\nu}  \right) \right. \\
&\left. +\,A\,{\bf v_e} \cdot {\bf \hat s_n} (1 - {\bf v_\nu} \cdot {\bf \hat s_\nu} )  - a\,\gamma_\nu^{-1}{\bf v_e} \cdot {\bf \hat s_\nu}    \right. \\
&\left.+\,a\,{\bf  v_e} \cdot {\bf  v_\nu} \left( 1 -{\gamma_\nu \over \gamma_\nu + 1}{ {\bf  v_\nu} } \cdot {\bf \hat s_\nu} \right) \right] \,,
}
where the asymmetry parameters are defined as
\es{asymm}{
a &= {1 - g_A^2 \over 1 + 3 g_A^2} \,, \,\,\, A = {2 g_A(1 - g_A) \over 1 + 3 g_A^2} \,, \,\,\,
 B = {2 g_A(1 + g_A) \over 1 + 3 g_A^2}  \,.
}
This scattering amplitude is one of the central results of this paper.  Averaging over the neutron polarization and restricting to neutrino helicity eigenstates,  \eqref{spinSum} agrees with the results in~\cite{Long:2014zva}.  
In the relativistic limit for the neutrino, the amplitude vanishes for right-handed neutrinos, with ${\bf v_\nu} \cdot {\bf \hat s_\nu} = 1$.  For left-handed relativistic neutrinos, with  ${\bf v_\nu} \cdot {\bf \hat s_\nu} = -1$, 
\es{}{
\overbar{ |{\cal M}|^2} &= 16 G_F^2 c_1^2 |U_{ej}|^2 m_n m_p E_\nu E_e (1 + 3 g_A^2)\times \\
& \left( 1+ B\,{\bf v_\nu \cdot \bf \hat s_n} + A\,{\bf v_e} \cdot {\bf \hat s_n} + a\,{\bf v_e} \cdot {\bf v_\nu}  \right) \,.
}
In this limit, the amplitude agrees with the well-known neutron beta-decay amplitude, which is related to this process by a crossing symmetry.  When the polarization of the neutron and the direction of the outgoing electron are not observed, the relativistic limit of the amplitude also agrees with previous NCB calculations (see, for example,~\cite{Cocco:2007za}).  

Tritium decay is similar to neutron decay.  The reason is that the transition from $^3$H to $^3$He is superallowed, and superallowed transitions are determined by the isospin quantum numbers of the initial and final states, to a good approximation.  Tritium and $^3$He form an isospin doublet, just like the neutron and the proton.

The NCB amplitude obeys~\eqref{spinSum}, taking \mbox{$n\rightarrow\,^3$H} and \mbox{$p\rightarrow\, ^3$He}, and making the appropriate kinematic substitutions.  However, when evaluating the asymmetry parameters~\eqref{asymm}, we should take (see, for example,~\cite{Grotz:1990jf,Schiavilla:1998je}) 
   \es{lambda}{
   g_A \to g_A  {\langle\text{\bf GT}  \rangle \over \sqrt{3} \langle {\text{\bf F} } \rangle } \approx 1.21 \,.
   }
 Above, $\langle\text{\bf GT}  \rangle$ and $\langle {\text{\bf F} } \rangle$ are the standard Gamow-Teller and Fermi matrix elements between the initial and final nuclear states.  For neutron decay, \mbox{$\langle {\text{\bf F} } \rangle = 1$} and \mbox{$\langle\text{\bf GT}  \rangle = \sqrt{3}$}.  For tritium decay~\cite{Schiavilla:1998je},
   \es{}{
   \langle {\text{\bf F} } \rangle^2 \approx 0.9987 \,, \qquad \langle {\text{\bf GT} } \rangle \approx \sqrt{3} \cdot 0.957 \,.
   }  
Note that the matrix element squared in~\eqref{spinSum} is multiplied by an overall factor of $\langle {\text{\bf F} } \rangle^2$, which we ignore as it is close to unity.  After making the substitution defined in~\eqref{lambda}, the asymmetry parameters evaluate to 
   \es{asymmP}{
   a \approx - 0.087 \,, \qquad A \approx -0.095 \,, \qquad B \approx 0.99 \,.
 }

The capture cross section can be calculated from the amplitude in~\eqref{spinSum}.  We define $\sigma_0$ to be the total $^3$H-polarization-averaged cross section.  For relic neutrinos with energies significantly below the beta-decay endpoint energy~\cite{Cocco:2007za,Long:2014zva}, 
  \es{sigma0}{
 \left. \sigma_0  v_\nu  \right|_{E_\nu = 0} &=(1 - {\bf \hat s_\nu } \cdot {\bf v_\nu} ) \, \bar \sigma \\
 \bar \sigma &=   |U_{ej}|^2 \times  3.83 \times 10^{-45} \, \, \text{cm}^2  \,.
  }
 
For directional detection of the C$\nu$B, we are interested in how the non-relativistic limit of the differential scattering amplitude depends on the direction of the $^3$H polarization, ${\bf \hat s_H}$, and the direction of the outgoing electron.
To first order in the neutrino velocity,
  \es{sDepS}{
  &{d\sigma({\bf \hat s_H}, {\bf \hat v_e} ) \over d \Omega_e} v_\nu \approx {\bar \sigma  \over 4 \pi} \left[ 1 - {\bf \hat s_\nu } \cdot {\bf v_\nu}  +B \,  {\bf \hat s_H} \cdot ({\bf v_\nu} - {\bf \hat s_\nu} )  \right. \\
  & \left. + \, A \, {\bf \hat s_H} \cdot {\bf v_e}\, (1 - {\bf \hat s_\nu } \cdot {\bf v_\nu} )+ a\, {\bf v_e} \cdot ({\bf v_\nu} - {\bf \hat s_\nu} )  \right] \,.
  } 
  
  \section{Applications to C$\nu$B detection}
  
Having derived the differential cross section for neutrino capture on a polarized tritium target, we now turn to its  implications for C$\nu$B detection.  We begin by discussing the total unpolarized capture rate
  \es{}{
  \bar \Gamma = N_\text{H} n_\nu \langle \sigma_0 v_\nu \rangle \,,
  }
where $N_\text{H}$ is the number of tritium nuclei in the detector and $n_\nu$ is the local neutrino number density.  Here, expectation values are taken both with respect to the relic neutrino phase-space distribution and with respect to the distribution of neutrino spins.  
  
The total capture rate depends on whether the neutrinos are Majorana or Dirac, as recently pointed out in~\cite{Long:2014zva}.  In the standard cosmology, equal populations of relativistic left- and right-handed active neutrinos decoupled from the thermal plasma.  If the neutrinos are Dirac, then the active right-handed states are antineutrinos.  Antineutrinos are not captured on tritium.  On the other hand, if the neutrinos are Majorana, then antineutrinos are indistinguishable from neutrinos.  The present-day relic neutrino number density is twice as large in the Majorana case than in the Dirac case.  The present-day number density for Dirac neutrinos is $n_{\nu} \approx 56 \, \text{cm}^{-3}$ per flavor, neglecting possible enhancements due to gravitational clustering. 

Let us begin by considering the relativistic limit for the incoming neutrinos.  
The cross section in~\eqref{sigma0} depends on \mbox{${\bf \hat s_\nu} \cdot {\bf v_\nu }$}.
In the Dirac scenario, all neutrinos are left-handed, $\langle {\bf \hat s_\nu} \cdot {\bf v_\nu }\rangle = -1$, while in the Majorana scenario, $\langle {\bf \hat s_\nu} \cdot {\bf v_\nu }\rangle = 0$.  
Thus, $\langle \sigma_0 v_\nu \rangle$ is twice as large for relic Dirac neutrinos, compared to Majorana neutrinos.  This factor of two compensates for the difference in the number densities.  As a result, the capture rates for relativistic neutrinos are the same in both the Majorana and Dirac scenarios.

In the non-relativistic limit, however, the unpolarized total cross section is independent of the neutrino spin.  As a result, the capture rate is twice as large in the Majorana case than in the Dirac one, due to the enhanced number density:
\es{}{
\bar \Gamma \approx  10 \,{ \text{events} \over \text{year} } \cdot { M_\text{Det} \over 100 \, \text{g}}\cdot {n_\nu \over 112 \, \text{cm}^3 } \cdot \sum_j|U_{ej}|^2  \,.
}
Above, $M_\text{Det}$ is the mass of tritium in the detector, and the sum over mass eigenstates $\nu_j$ is over all states with masses above the detector threshold.  In practice, it will be difficult to identify whether such an enhancement in the rate is due to the Majorana versus Dirac nature of the neutrinos, or from an over-density due to gravitational clustering.  A better understanding of the local neutrino velocity distribution, obtained from annual modulation~\cite{Safdi:2014rza} or from measurements of the polarized differential rate, discussed below, could disambiguate the cause of a rate enhancement.  

Now we discuss the polarized differential rate 
\es{diffR}{
\frac{d \Gamma({\bf \hat s_H}, {\bf \hat v_e} )}{d \Omega_e} = N_\text{H} n_\nu \left\langle  {d\sigma({\bf \hat s_H}, {\bf \hat v_e} ) \over d \Omega_e} v_\nu \right\rangle \,. 
}
There are two types of terms that appear in~\eqref{diffR}: those proportional to $\langle 1 - {\bf \hat s_\nu } \cdot {\bf v_\nu} \rangle$ and those proportional to $\langle{\bf v_\nu} - {\bf \hat s_\nu} \rangle$.  To leading order in the non-relativistic limit, $\langle 1 - {\bf \hat s_\nu } \cdot {\bf v_\nu} \rangle$ is simply unity.  However, evaluating $\langle{\bf v_\nu} - {\bf \hat s_\nu} \rangle$ in this limit requires some care.

\begin{figure}[tb]
\begin{center}
\includegraphics[width=3.4in]{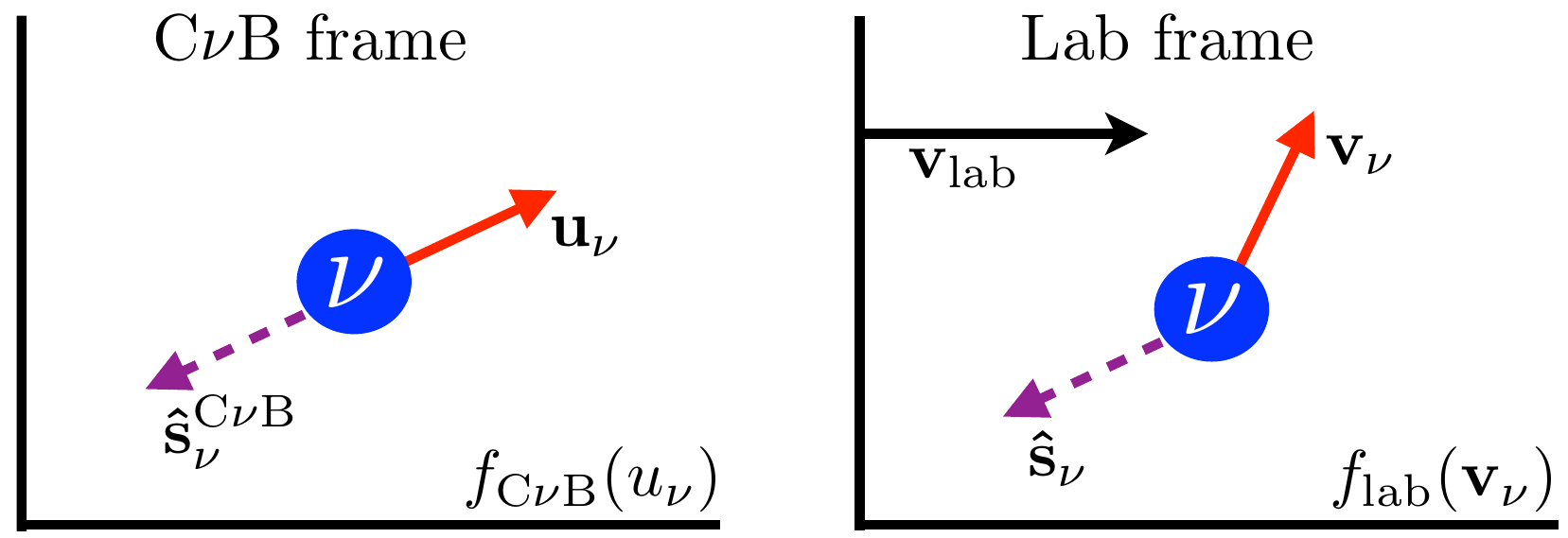}
\end{center}
\vspace{-.50cm}
\caption{
The velocity distribution $f_\text{C$\nu$B}(u_\nu)$ is isotropic in the C$\nu$B frame, with ${\bf u_\nu}$ the neutrino velocity (red, solid) in that frame.  In this illustration, the neutrino is left-handed in the C$\nu$B frame, so its spin (purple, dotted) is ${\bf \hat s_\nu^\text{C$\nu$B} }= - {\bf \hat u_\nu}$.  The lab frame is boosted with respect to the C$\nu$B frame by ${\bf v_\text{C$\nu$B}}$, and so the lab-frame neutrino velocity is ${\bf v_\nu }= {\bf u_\nu} - {\bf v_\text{C$\nu$B}}$.  
  }      
\vspace{-0.15in}  
\label{Frames}
\end{figure}

Let $\textbf{u}_\nu$ ($\textbf{v}_\nu$) and ${\bf \hat s_\nu^\text{C$\nu$B}}$ (${\bf \hat s_\nu}$)  be the velocity and spin of the neutrino in the C$\nu$B (lab) frame, respectively.  The lab-frame neutrino velocities ${\bf v_\nu}$ are related to ${\bf u_\nu}$ through a Galilean transformation: ${\bf u_\nu} = {\bf v_\nu} + {\bf v_\text{lab}}$, where $ {\bf v_\text{lab}}$ is the velocity of the lab in the C$\nu$B frame.  (See Fig.~\ref{Frames} for an illustration.)  We assume that the normalized neutrino velocity distribution $f_\text{C$\nu$B}$ is isotropic in the C$\nu$B rest frame; $f_\text{C$\nu$B}( {\bf u_\nu} ) = f_\text{C$\nu$B}( u_\nu ) $.  In particular, $f_\text{lab}({\bf v_\nu}) = f_\text{C$\nu$B}( |{\bf v_\nu} + {\bf v_\text{lab}}|)$.  The average velocity of the neutrinos in the lab frame is then
  \es{}{
 \langle {\bf v_\nu} \rangle = \int d^3 v_\nu \, f_\text{lab}({\bf v_\nu}) \, {\bf v_\nu } = - {\bf v_\text{lab}} \,.
  }

What remains is to calculate  the expectation value of the neutrino spin vector in the lab frame, $\langle {\bf \hat s_\nu}\rangle$.  The neutrinos are in helicity eigenstates at decoupling and, neglecting gravitational effects, they remain in those states today (in the cosmic frame) because the helicity operator is conserved.  However, because the neutrinos are non-relativistic, their helicities are affected when boosting to the lab frame.  To address this issue, we need to examine how a spin three-vector changes under the change of frames.  The product of the boost from the neutrino's rest frame, where its spin is defined, to the C$\nu$B frame and the boost from the C$\nu$B frame to the lab frame may be decomposed into a single boost times a rotation.  The spin three-vector is invariant under the boost, but it rotates by the well-known Wigner angle under the rotation (see~\cite{BenMenahem} for a review).  In the non-relativistic limit, the neutrino's lab-frame spin is therefore
\es{Wigner}{
{\bf \hat s_\nu} = {\bf \hat s_\nu^\text{C$\nu$B}} + \frac{1}{2} u_\nu v_\text{lab} \left( {\bf \hat u_\nu} {\bf \hat v_\text{lab}} - {\bf \hat v_\text{lab}} {\bf \hat u_\nu} \right)\cdot {\bf \hat s_\nu^\text{C$\nu$B}} \,,
}  
to leading-order in the boost velocities.  The second term in \eqref{Wigner} is suppressed by two factors of the speed of light, so to first approximation, ${\bf \hat s_\nu} \approx {\bf \hat s_\nu^\text{C$\nu$B}}$.  

If the neutrinos are Majorana, then they are equally likely to be left- or right-handed in the cosmic frame.  In this case,
\es{WignerM}{
\langle{\bf \hat s_\nu}\rangle_\text{M} \approx \langle {\bf \hat s_\nu^\text{C$\nu$B}}\rangle_\text{M} = 0 \, .
}  
If the neutrinos are Dirac, then they are purely left-handed in the cosmic frame, so their spin is oriented opposite the direction of motion.  In this case, 
\es{WignerD}{
\langle{\bf \hat s_\nu}\rangle_\text{D} \approx \langle {\bf \hat s_\nu^\text{C$\nu$B}}\rangle_\text{D} = - \langle {\bf \hat u_\nu} \rangle = 0 \, 
}  
because the cosmic-frame velocities are isotropic.  Therefore, the term proportional to $\langle {\bf \hat s_\nu} \rangle$ in \eqref{diffR} can be safely ignored for both the Majorana and Dirac cases.

The differential rate depends on the electron's velocity through the terms 
\es{electronV}{
\frac{d \Gamma({\bf \hat s_H}, {\bf \hat v_e} )}{d \Omega_e}  \supset  {N_H n_\nu  \bar \sigma  \over 4 \pi}  \left(  A \, {\bf \hat s_H} \cdot {{\bf v_e} \over c} - a\, {{\bf v_e} \over c}\cdot {{\bf v_\text{lab}} \over c} \right) \,.
 } 
  Note that we have re-instated the speed of light $c$.
The second term in \eqref{electronV} is subdominant compared to the first as it is proportional to $v_\text{lab} / c$.  From the `$A$' asymmetry term, we see that the electrons tend to be emitted away from the direction of polarization.  If this asymmetry can be measured, then it is convincing evidence that the electrons are coming from the tritium and not from some other background source.  However, the beta-decay electrons have the same preference to be emitted away from the polarization direction, so one cannot use this asymmetry parameter to distinguish between NCB and beta decay.  With further exposure, it may be possible to measure the `$a$' asymmetry term.  Because $a$ is negative, this asymmetry is manifested by a slight preference for the electrons to be emitted in the direction ${\bf \hat v_\text{lab}}$.    

Even without observing the direction of the outgoing electron, we may extract directional information by studying the total polarized rate:
\es{GammaSH0}{
\Gamma({\bf \hat s_H}) &= \int d \Omega_e \,\frac{d \Gamma({\bf \hat s_H}, {\bf \hat v_e} )}{d \Omega_e} \\ 
& = N_\text{H} \, n_\nu \, \bar \sigma \left(1 - B {{\bf v}_\text{lab} \over c} \cdot {\bf \hat s_H} \right) \,.
}
The capture rate is maximal when the tritium polarization is anti-aligned with the lab-frame velocity and minimal when the two vectors are aligned.  If a target on Earth is prepared with a particular polarization, the angle between ${\bf \hat s_H}$ and ${\bf v_\text{lab}}$ will change during the day as the Earth rotates, resulting in a daily-modulating rate.

The modulation fraction depends on the lab-frame velocity ${\bf v_\text{lab}}$, which in turn depends on the clustering of the cosmic neutrinos.  The local phase-space distribution of relic neutrinos is not well understood and requires careful numerical simulations.  Here, we follow~\cite{Safdi:2014rza} and assume two limiting cases for illustration.  If the relic neutrinos are bound and isotropic in the Galactic rest frame, then \mbox{${\bf v}_\text{lab} = {\bf v}_\text{MW}   \approx 232 \, (0.047 ,0.998 ,0.030 ) \kms$} is the velocity of the Sun with respect to the Galactic Center in Galactic coordinates~\cite{Schoenrich:2009bx}.  In the opposite limit, the relic neutrinos are unperturbed by the Milky Way.  Then, the C$\nu$B frame is the same as the CMB frame~\cite{Dodelson:2009ze}.  The Sun travels with velocity \mbox{${\bf v}_\text{lab} = {\bf v}_\text{CMB} \approx 369 \, (-0.0695, -0.662, 0.747) \kms$} with respect to the CMB rest frame~\cite{Kogut:1993ag,Hinshaw:2008kr,Aghanim:2013suk}.  In either of these limiting cases, the modulation fraction is 
\es{modFrac}{
\mathcal{O}\left(B \, \frac{v_\text{lab}}{c}\right)\sim0.1\% \, .
}

\begin{figure}[tb]
\begin{center}
\includegraphics[width=3.4in]{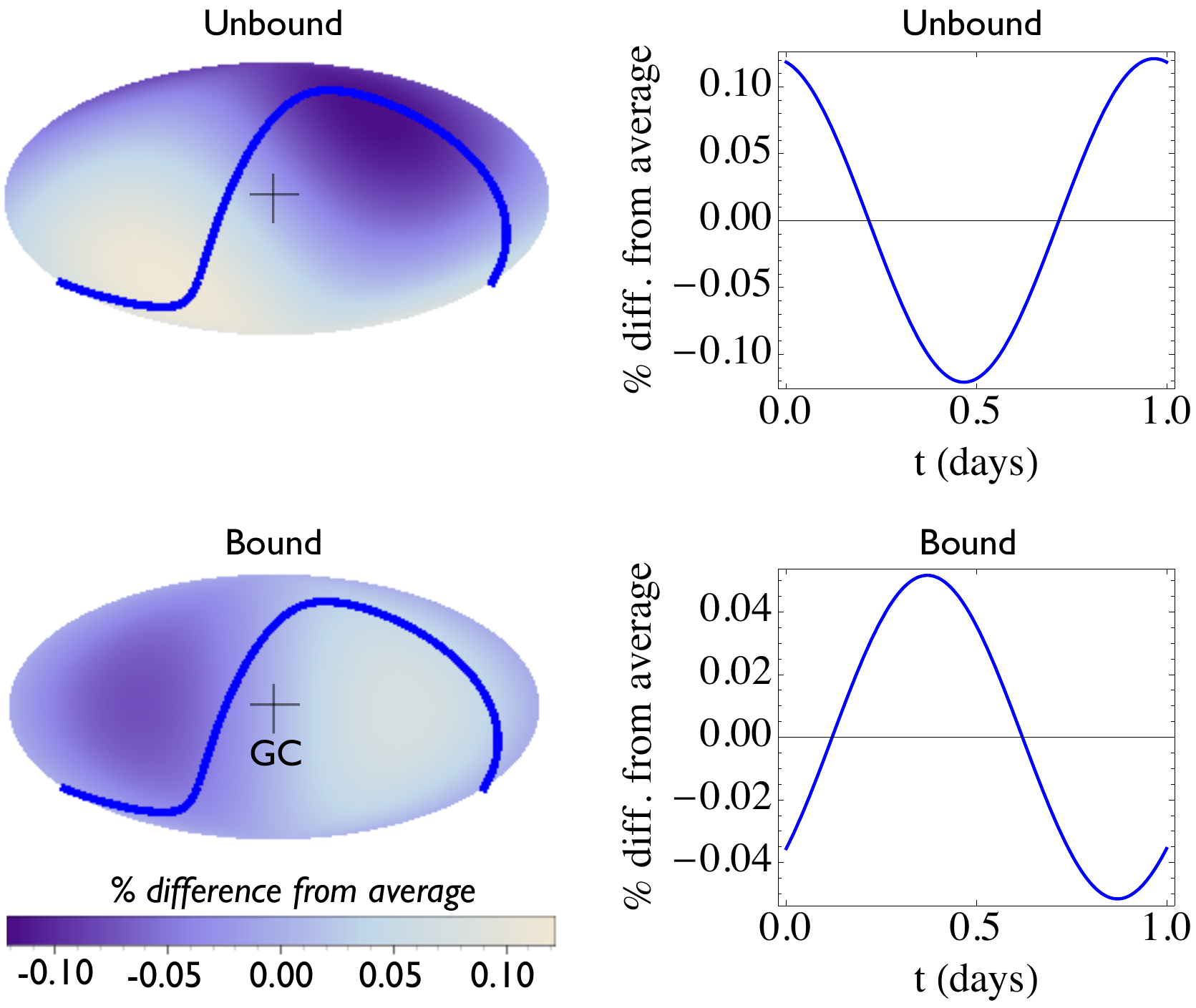}
\end{center}
\vspace{-.50cm}
\caption{
The NCB detection rate for the C$\nu$B depends on the direction of the polarization vector ${\bf \hat s_H}$, shown in the left column in Galactic coordinates (Mollweide projection), with the Galactic Center (GC) at the origin.  
A polarization vector ${\bf \hat s_H}$ that is fixed on the surface of the Earth sweeps out a circle in the sky during the day, and is manifested in terms of a daily modulation as a result of the dipole asymmetry.  We illustrate this for a polarization vector aligned perpendicular to the Earth's rotational axis.  The path through the sky is shown by the solid blue curves through the Mollweide maps, and the daily-modulating rates are shown in the right column.  
In the bound(unbound) map, the direction of minimal rate corresponds to the direction ${\bf \hat v}_\text{MW}$(${\bf \hat v}_\text{CMB}$).
  }      
\vspace{-0.15in}  
\label{Fig: P1}
\end{figure}

Figure~\ref{Fig: P1} illustrates the dependence of the capture rate on the direction of the polarization vector ${\bf \hat s_H}$ for the bound and unbound scenarios.  The example in Fig.~\ref{Fig: P1} has the polarization vector aligned perpendicular to the Earth's rotational axis.  A fixed polarization direction on Earth sweeps out circles in the sky during the course of a day, assuming ${\bf \hat s_H}$ is not aligned with the Earth's rotational axis.  The daily trajectory of ${\bf \hat s_H}$ through the sky is shown by the solid curve superimposed on the Mollweide projections (left column).  

In the Mollweide maps, the directions of minimal(maximal) rate correspond to the directions ${\bf \hat v_\text{lab}}$($-{\bf \hat v_\text{lab}}$).  The difference between ${\bf \hat v_\text{lab}}$ in the unbound and bound scenarios accounts for the difference in their respective daily modulation phases.  The bound scenario has an additional suppression in Fig.~\ref{Fig: P1} relative to the unbound scenario because in that example the vector ${\bf \hat s_H}$ is never aligned with ${\bf \hat v}_\text{MW}$ during the course of the day while it is aligned with ${\bf \hat v}_\text{CMB}$.  

The C$\nu$B is expected to be nearly isotropic at Earth's location, with a small dipole anisotropy suppressed by the lab-frame speed divided by $c$.  This is in analogy with the CMB dipole anisotropy~\cite{Kogut:1993ag,Hinshaw:2008kr,Aghanim:2013suk}.  The small C$\nu$B anisotropy is a non-trivial prediction of the thermal cosmology.  An experiment with $\sim$100 g of tritium, such as PTOLEMY, will not have enough exposure to observe the dipole anisotropy, assuming the neutrino over-density is not too significant.  

In the example we considered, the daily modulation fraction is suppressed by the factor $v_\text{lab}/c$.  The modulation can be more significant for non-standard scenarios where either $\langle{\bf \hat s_\nu}\rangle$ or $\langle {\bf v_\nu}\rangle$ is enhanced. 
For example, an anisotropic spin distribution in the lab frame could occur if the neutrino has a sufficiently large magnetic dipole moment such that helicity eigenstates become mixed while propagating through the Galactic magnetic fields.  The amount of time that the neutrinos have been subjected to these fields and the amplitudes of the fields depend on the arrival directions of the neutrinos at the Sun.  The neutrinos that have spent more time traversing regions of large fields will have mixed helicities, while those that have spent little time in the fields will remain in pure helicity states.  Thus, the fraction of left-handed neutrinos that have rotated into right-handed neutrinos depends on the neutrinos' directions.  This effect is only observable in the Dirac scenario, because if the neutrinos are Majorana, then there is an initial distribution of right-handed neutrinos that also rotate into left-handed neutrinos.
We leave a careful study of the phenomenology of relic neutrinos with magnetic dipole moments at polarized NCB observatories to future work.

\section{Feasibility
}

In the previous section, we showed that a polarized target can be used to probe anisotropies in the C$\nu$B.  Here, we briefly discuss the feasibility of this proposal, focusing specifically on the PTOLEMY experiment.  

As planned, the target at PTOLEMY will consist of $^3$H  that is atomically bound to graphene.  At low temperatures $T_\text{H}$ and in the presence of a strong external magnetic field $B$, the $^3$H nuclear spins align thermally due to the $^3$H magnetic dipole moment.  The thermal polarization fraction is easily estimated to be \mbox{$P = \tanh (\alpha /  2)$}, with \mbox{$\alpha \approx 0.02 (B / 10 \, \text{T}) (1 \, \text{K} / T_\text{H})$}.\footnote{It has recently been shown that certain hydrogenations of graphene exhibit ferromagnetism~\cite{doi:10.1021/nl9020733,doi:10.1021/nn4016289}.  This may help align the $^3$H nuclear spins by increasing the internal magnetic field.}
However, there are multiple dynamical polarization techniques that may be applied to the atomically-held tritium system in order to achieve polarization fractions significantly above the thermal estimate.  For example, it may be possible to use the Overhauser effect~\cite{PhysRev.92.411,1962JChPh..37...85A} for certain hydrogenations of the graphene that are semiconducting~\cite{PhysRevLett.92.225502,PhysRevB.77.035427,Balog:2010fk}, such as the same-sided fully-hydrogenated graphene~\cite{PhysRevB.84.041402}.    This method involves transferring the polarization of unpaired electrons to the atomic nuclei through microwave pumping. 
Further study is necessary to determine the optimal mechanism for polarizing the tritium nuclei under the conditions planned for PTOLEMY.  Importantly, large polarization fractions must be maintained over an extended time period  to measure the `$A$' and `$B$' anisotropies in~\eqref{sDepS}.

The polarization fraction can be measured using nuclear magnetic resonance.  However, the tritiated graphene also provides a novel approach for studying the nuclear polarization of hydrogenated graphene.  The $^3$H beta-decay electrons exhibit an asymmetry with respect to the polarization axis, captured by the electron asymmetry parameter `$A$' in~\eqref{spinSum}; the electrons tend to be emitted away from the polarization axis because `$A$' is negative. 
 By measuring the asymmetry of the beta-decay electrons with respect to the direction of the external magnetic field, one may infer the polarization fraction of the material. 

Measuring the `$A$' and `$a$' anisotropies in the differential rate requires sensitivity to the electron's  velocity.  PTOLEMY should be able to measure the projection of this velocity perpendicular to the direction of the solenoid.  This is accomplished by tracking the RF signal from the cyclotron motion and through time-of-flight measurements~\cite{Betts:2013uya}.
The `A' and `a' asymmetries are easily separated by studying the evolution of the total electron asymmetry throughout the course of the day.  That is, the direction of the `a' asymmetry modulates throughout the day, due to the change in the lab-frame orientation, while the `A' asymmetry is static, since ${\bf \hat v_\text{H}}$ is static in the lab frame.  Measuring the `A' asymmetry requires changing the angle between the tritium polarization and the solenoid magnetic field.

Measuring the asymmetries of the neutrino capture cross section is a fundamentally new approach to C$\nu$B studies.  To ensure that the experimental and theoretical uncertainties are under control, one would first want to calibrate the detector by studying the relativistic limit of the polarization-dependent NCB cross section.  This can be done by placing an external neutrino source near the detector.  As an example, we consider a $^{51}$Cr neutrino source placed a distance $D$ from the detector with an activity $\Gamma_\text{source}$.  The isotope $^{51}$Cr decays via electron capture to $^{51}$V.  The emitted neutrinos are mono-energetic, at energies $746$ keV (81\%), 751 keV (9\%), 426 keV (9\%), and 431 keV (1\%)~\cite{Hampel:1997fc,Abdurashitov:1998ne}. 
$^{51}$Cr neutrino sources with $\Gamma_\text{source} \sim {\cal O}(\text{MCi})$ have been used successfully at the GALLEX~\cite{Hampel:1997fc} and SAGE~\cite{Abdurashitov:1998ne} experiments in the past.  The BOREXINO collaboration~\cite{Borexino:2013xxa} has also discussed using artificial $^{51}$Cr neutrinos to study short-distance neutrino oscillation.  

The neutrinos produced in the decay of $^{51}$Cr are much more energetic than relic neutrinos, which means that the relevant NCB cross section is enhanced over the low-energy cross section.   In this regime,
\es{sigmaE}{
\sigma_0 v_\nu \approx \left. \sigma_0 v_\nu \right|_{E_\nu = 0}  {E_e \over m_e + Q_\beta} {p_e \over \sqrt{2 m_e Q_\beta}}{ F(E_e) \over F(m_e + Q_\beta)} \,,
}
where $F(E_e)$ is the Fermi function for $^3$He, \mbox{$Q_\beta \approx 18.6$ keV} is the beta-decay endpoint energy, and $m_e$ is the electron mass.  It follows that the polarization-averaged detection rate of $^{51}$Cr neutrinos at a $^{3}$H NCB experiment is 
\es{}{
\Gamma^{^{51}\text{Cr}} \approx 4 \times 10^3 \,{ \text{events} \over \text{year} } \, { M_\text{Det.} \over 100 \, \text{g}} \, {\Gamma_\text{source} \over 100 \, \text{MCi} } \left( {1 \, \text{m} \over \text{D}} \right)^2 \,.
}
Clearly, an experiment such as PTOLEMY would see a substantial number of events from the decay of the $^{51}$Cr source.  This signal should modulate as the polarization direction is rotated in and out of alignment with the neutrino beam.  For these relativistic neutrinos, the modulation of the total rate is set by the term $B \, {\bf \hat v_\nu}\cdot{\bf \hat s_H}$, resulting in a $\sim$100\% modulation fraction.  It may also be possible to use such a setup to search for new neutrino physics, such as $\mathcal{O}(\text{eV})$ sterile neutrinos with small mixing to the active neutrino eigenstates.
\newline

\section{Conclusions} 

We presented a novel method for probing the dipole anisotropies in the spin and velocity distributions of the cosmic neutrino background.  The neutrino capture rate depends on the angular separation of the polarization axis of the nucleus with the neutrino's momentum, as well as its angular separation with the neutrino's spin.  For NCB on a tritium target, the neutrinos are preferentially captured when they approach along the polarization axis and when their spins are anti-aligned with the polarization axis.  Similar anisotropies exist for the differential capture rate as a function of the direction of the outgoing electron.

Our proposal is of relevance for the PTOLEMY experiment, which plans to use a $\sim$100 g $^3$H target atomically bound to graphene to detect the C$\nu$B.  The C$\nu$B should have a small dipole anisotropy, of order $\sim$0.1\%.  Therefore, measuring a nearly isotropic distribution of low-energy neutrinos would serve as a strong test of the cosmological origin of a potential signal.

The dipole anisotropy is directly related to the average velocity of the lab frame with respect to the C$\nu$B.  Annual modulation of relic neutrinos may allow for additional characterization of the neutrino background~\cite{Safdi:2014rza}.  If one already knows the average relative velocity between the lab and cosmic frames, then the amplitude of an annually-modulating signal would directly probe the velocity dispersion of the C$\nu$B.  Thus, a combination of these two measurements can be used to infer the temperature of the C$\nu$B.  Additionally, since the polarized-target and modulation measurements both characterize the relic neutrino velocity distribution, these observations may directly constrain the fraction of bound versus unbound neutrinos.  It is then possible to determine whether the neutrinos are non-relativistic Dirac or Majorana, since the capture rate is twice as large in the latter scenario; the phase-space probes break the degeneracy between Dirac versus Majorana and a local C$\nu$B over-density. 

The asymmetries in the neutrino capture cross section only allow for a measurement of the C$\nu$B dipole asymmetry.
 It is important to eventually characterize the higher multipole moments.  Towards that end, it would be useful to find methods for improving the angular resolution of ultra-low energy neutrino measurements.

\vspace{0.1in}
\noindent{\it Note added:  Ref.~\cite{Long:2014zva}, which appeared as this work was being completed, studies the physics potential of $C\nu B$ detection.  The scattering amplitude that we calculate agrees with theirs when averaged over the neutron spin and restricted to neutrino helicity eigenstates.}

\vspace{-0.3in}
\section*{Acknowledgments}
\noindent We thank F.~Froborg, R.~D'Agnolo, W. Happer, S.~Lee, S.~Pufu, and J.~Suerfu for helpful discussions.  BRS is supported by the NSF grant PHY-1314198.  CGT is supported in part by the DOE Award \# ER-41850.

\vspace{0in}
\onecolumngrid
\vspace{0.3in}
\twocolumngrid
\def\bibsection{} 
\bibliographystyle{apsrev}
\bibliography{polarizedNeutrinos}

\begin{thebibliography}{30}
\expandafter\ifx\csname natexlab\endcsname\relax\def\natexlab#1{#1}\fi
\expandafter\ifx\csname bibnamefont\endcsname\relax
  \def\bibnamefont#1{#1}\fi
\expandafter\ifx\csname bibfnamefont\endcsname\relax
  \def\bibfnamefont#1{#1}\fi
\expandafter\ifx\csname citenamefont\endcsname\relax
  \def\citenamefont#1{#1}\fi
\expandafter\ifx\csname url\endcsname\relax
  \def\url#1{\texttt{#1}}\fi
\expandafter\ifx\csname urlprefix\endcsname\relax\def\urlprefix{URL }\fi
\providecommand{\bibinfo}[2]{#2}
\providecommand{\eprint}[2][]{\url{#2}}

\bibitem[{\citenamefont{Weinberg}(2008)}]{Weinberg:2008zzc}
\bibinfo{author}{\bibfnamefont{S.}~\bibnamefont{Weinberg}},
  \emph{\bibinfo{title}{{Cosmology}}} (\bibinfo{publisher}{OUP},
  \bibinfo{address}{New York}, \bibinfo{year}{2008}).

\bibitem[{\citenamefont{Dicke et~al.}(1965)\citenamefont{Dicke, Peebles, Roll,
  and Wilkinson}}]{Dicke:1965zz}
\bibinfo{author}{\bibfnamefont{R.}~\bibnamefont{Dicke}},
  \bibinfo{author}{\bibfnamefont{P.}~\bibnamefont{Peebles}},
  \bibinfo{author}{\bibfnamefont{P.}~\bibnamefont{Roll}}, \bibnamefont{and}
  \bibinfo{author}{\bibfnamefont{D.}~\bibnamefont{Wilkinson}},
  \bibinfo{journal}{Astrophys.J.} \textbf{\bibinfo{volume}{142}},
  \bibinfo{pages}{414} (\bibinfo{year}{1965}).

\bibitem[{\citenamefont{Weinberg}(1962)}]{Weinberg:1962zza}
\bibinfo{author}{\bibfnamefont{S.}~\bibnamefont{Weinberg}},
  \bibinfo{journal}{Phys.Rev.} \textbf{\bibinfo{volume}{128}},
  \bibinfo{pages}{1457} (\bibinfo{year}{1962}).

\bibitem[{\citenamefont{Ade et~al.}(2013)}]{Ade:2013zuv}
\bibinfo{author}{\bibfnamefont{P.}~\bibnamefont{Ade}} \bibnamefont{et~al.}
  (\bibinfo{collaboration}{Planck Collaboration}) (\bibinfo{year}{2013}),
  \eprint{1303.5076}.

\bibitem[{\citenamefont{Beringer et~al.}(2012)}]{Beringer:1900zz}
\bibinfo{author}{\bibfnamefont{J.}~\bibnamefont{Beringer}} \bibnamefont{et~al.}
  (\bibinfo{collaboration}{Particle Data Group}), \bibinfo{journal}{Phys.Rev.}
  \textbf{\bibinfo{volume}{D86}}, \bibinfo{pages}{010001}
  (\bibinfo{year}{2012}).

\bibitem[{\citenamefont{Betts et~al.}(2013)\citenamefont{Betts, Blanchard,
  Carnevale, Chang, Chen et~al.}}]{Betts:2013uya}
\bibinfo{author}{\bibfnamefont{S.}~\bibnamefont{Betts}},
  \bibinfo{author}{\bibfnamefont{W.}~\bibnamefont{Blanchard}},
  \bibinfo{author}{\bibfnamefont{R.}~\bibnamefont{Carnevale}},
  \bibinfo{author}{\bibfnamefont{C.}~\bibnamefont{Chang}},
  \bibinfo{author}{\bibfnamefont{C.}~\bibnamefont{Chen}}, \bibnamefont{et~al.}
  (\bibinfo{year}{2013}), \eprint{1307.4738}.

\bibitem[{\citenamefont{Long et~al.}(2014)\citenamefont{Long, Lunardini, and
  Sabancilar}}]{Long:2014zva}
\bibinfo{author}{\bibfnamefont{A.~J.} \bibnamefont{Long}},
  \bibinfo{author}{\bibfnamefont{C.}~\bibnamefont{Lunardini}},
  \bibnamefont{and}
  \bibinfo{author}{\bibfnamefont{E.}~\bibnamefont{Sabancilar}}
  (\bibinfo{year}{2014}), \eprint{1405.7654}.

\bibitem[{\citenamefont{Ringwald and Wong}(2004)}]{Ringwald:2004np}
\bibinfo{author}{\bibfnamefont{A.}~\bibnamefont{Ringwald}} \bibnamefont{and}
  \bibinfo{author}{\bibfnamefont{Y.~Y.} \bibnamefont{Wong}},
  \bibinfo{journal}{JCAP} \textbf{\bibinfo{volume}{0412}}, \bibinfo{pages}{005}
  (\bibinfo{year}{2004}), \eprint{hep-ph/0408241}.

\bibitem[{\citenamefont{Lee et~al.}(2014)\citenamefont{Lee, Lisanti, Peter, and
  Safdi}}]{Lee:2013wza}
\bibinfo{author}{\bibfnamefont{S.~K.} \bibnamefont{Lee}},
  \bibinfo{author}{\bibfnamefont{M.}~\bibnamefont{Lisanti}},
  \bibinfo{author}{\bibfnamefont{A.~H.~G.} \bibnamefont{Peter}},
  \bibnamefont{and} \bibinfo{author}{\bibfnamefont{B.~R.} \bibnamefont{Safdi}},
  \bibinfo{journal}{Phys.Rev.Lett.} \textbf{\bibinfo{volume}{112}},
  \bibinfo{pages}{011301} (\bibinfo{year}{2014}), \eprint{1308.1953}.

\bibitem[{\citenamefont{Safdi et~al.}(2014)\citenamefont{Safdi, Lisanti, Spitz,
  and Formaggio}}]{Safdi:2014rza}
\bibinfo{author}{\bibfnamefont{B.~R.} \bibnamefont{Safdi}},
  \bibinfo{author}{\bibfnamefont{M.}~\bibnamefont{Lisanti}},
  \bibinfo{author}{\bibfnamefont{J.}~\bibnamefont{Spitz}}, \bibnamefont{and}
  \bibinfo{author}{\bibfnamefont{J.~A.} \bibnamefont{Formaggio}}
  (\bibinfo{year}{2014}), \eprint{1404.0680}.

\bibitem[{\citenamefont{Cocco et~al.}(2007)\citenamefont{Cocco, Mangano, and
  Messina}}]{Cocco:2007za}
\bibinfo{author}{\bibfnamefont{A.~G.} \bibnamefont{Cocco}},
  \bibinfo{author}{\bibfnamefont{G.}~\bibnamefont{Mangano}}, \bibnamefont{and}
  \bibinfo{author}{\bibfnamefont{M.}~\bibnamefont{Messina}},
  \bibinfo{journal}{JCAP} \textbf{\bibinfo{volume}{0706}}, \bibinfo{pages}{015}
  (\bibinfo{year}{2007}), \eprint{hep-ph/0703075}.

\bibitem[{\citenamefont{Grotz and Klapdor}(1990)}]{Grotz:1990jf}
\bibinfo{author}{\bibfnamefont{K.}~\bibnamefont{Grotz}} \bibnamefont{and}
  \bibinfo{author}{\bibfnamefont{H.}~\bibnamefont{Klapdor}},
  \emph{\bibinfo{title}{{The Weak interaction in nuclear, particle and
  astrophysics}}} (\bibinfo{publisher}{CRC Press}, \bibinfo{year}{1990}).

\bibitem[{\citenamefont{Schiavilla et~al.}(1998)\citenamefont{Schiavilla,
  Stoks, Gloeckle, Kamada, Nogga et~al.}}]{Schiavilla:1998je}
\bibinfo{author}{\bibfnamefont{R.}~\bibnamefont{Schiavilla}},
  \bibinfo{author}{\bibfnamefont{V.}~\bibnamefont{Stoks}},
  \bibinfo{author}{\bibfnamefont{W.}~\bibnamefont{Gloeckle}},
  \bibinfo{author}{\bibfnamefont{H.}~\bibnamefont{Kamada}},
  \bibinfo{author}{\bibfnamefont{A.}~\bibnamefont{Nogga}},
  \bibnamefont{et~al.}, \bibinfo{journal}{Phys.Rev.}
  \textbf{\bibinfo{volume}{C58}}, \bibinfo{pages}{1263} (\bibinfo{year}{1998}),
  \eprint{nucl-th/9808010}.

\bibitem[{\citenamefont{BenMenahem}(1985)}]{BenMenahem}
\bibinfo{author}{\bibfnamefont{A.}~\bibnamefont{BenMenahem}},
  \bibinfo{journal}{Am. J. Phys.} \textbf{\bibinfo{volume}{53}},
  \bibinfo{pages}{62} (\bibinfo{year}{1985}).

\bibitem[{\citenamefont{Schoenrich et~al.}(2009)\citenamefont{Schoenrich,
  Binney, and Dehnen}}]{Schoenrich:2009bx}
\bibinfo{author}{\bibfnamefont{R.}~\bibnamefont{Schoenrich}},
  \bibinfo{author}{\bibfnamefont{J.}~\bibnamefont{Binney}}, \bibnamefont{and}
  \bibinfo{author}{\bibfnamefont{W.}~\bibnamefont{Dehnen}}
  (\bibinfo{year}{2009}), \eprint{0912.3693}.

\bibitem[{\citenamefont{Dodelson and Vesterinen}(2009)}]{Dodelson:2009ze}
\bibinfo{author}{\bibfnamefont{S.}~\bibnamefont{Dodelson}} \bibnamefont{and}
  \bibinfo{author}{\bibfnamefont{M.}~\bibnamefont{Vesterinen}},
  \bibinfo{journal}{Phys.Rev.Lett.} \textbf{\bibinfo{volume}{103}},
  \bibinfo{pages}{171301} (\bibinfo{year}{2009}), \eprint{0907.2887}.

\bibitem[{\citenamefont{Kogut et~al.}(1993)\citenamefont{Kogut, Lineweaver,
  Smoot, Bennett, Banday et~al.}}]{Kogut:1993ag}
\bibinfo{author}{\bibfnamefont{A.}~\bibnamefont{Kogut}},
  \bibinfo{author}{\bibfnamefont{C.}~\bibnamefont{Lineweaver}},
  \bibinfo{author}{\bibfnamefont{G.~F.} \bibnamefont{Smoot}},
  \bibinfo{author}{\bibfnamefont{C.}~\bibnamefont{Bennett}},
  \bibinfo{author}{\bibfnamefont{A.}~\bibnamefont{Banday}},
  \bibnamefont{et~al.}, \bibinfo{journal}{Astrophys.J.}
  \textbf{\bibinfo{volume}{419}}, \bibinfo{pages}{1} (\bibinfo{year}{1993}),
  \eprint{astro-ph/9312056}.

\bibitem[{\citenamefont{Hinshaw et~al.}(2009)}]{Hinshaw:2008kr}
\bibinfo{author}{\bibfnamefont{G.}~\bibnamefont{Hinshaw}} \bibnamefont{et~al.}
  (\bibinfo{collaboration}{WMAP Collaboration}),
  \bibinfo{journal}{Astrophys.J.Suppl.} \textbf{\bibinfo{volume}{180}},
  \bibinfo{pages}{225} (\bibinfo{year}{2009}), \eprint{0803.0732}.

\bibitem[{\citenamefont{Aghanim et~al.}(2013)}]{Aghanim:2013suk}
\bibinfo{author}{\bibfnamefont{N.}~\bibnamefont{Aghanim}} \bibnamefont{et~al.}
  (\bibinfo{collaboration}{Planck Collaboration}) (\bibinfo{year}{2013}),
  \eprint{1303.5087}.

\bibitem[{\citenamefont{Zhou et~al.}(2009)\citenamefont{Zhou, Wang, Sun, Chen,
  Kawazoe, and Jena}}]{doi:10.1021/nl9020733}
\bibinfo{author}{\bibfnamefont{J.}~\bibnamefont{Zhou}},
  \bibinfo{author}{\bibfnamefont{Q.}~\bibnamefont{Wang}},
  \bibinfo{author}{\bibfnamefont{Q.}~\bibnamefont{Sun}},
  \bibinfo{author}{\bibfnamefont{X.~S.} \bibnamefont{Chen}},
  \bibinfo{author}{\bibfnamefont{Y.}~\bibnamefont{Kawazoe}}, \bibnamefont{and}
  \bibinfo{author}{\bibfnamefont{P.}~\bibnamefont{Jena}},
  \bibinfo{journal}{Nano Letters} \textbf{\bibinfo{volume}{9}},
  \bibinfo{pages}{3867} (\bibinfo{year}{2009}).

\bibitem[{\citenamefont{Eng et~al.}(2013)\citenamefont{Eng, Poh, {\v S}an{\v
  e}k, Mary{\v s}ko, Mat{\v e}jkov{\'a}, Sofer, and
  Pumera}}]{doi:10.1021/nn4016289}
\bibinfo{author}{\bibfnamefont{A.~Y.~S.} \bibnamefont{Eng}},
  \bibinfo{author}{\bibfnamefont{H.~L.} \bibnamefont{Poh}},
  \bibinfo{author}{\bibfnamefont{F.}~\bibnamefont{{\v S}an{\v e}k}},
  \bibinfo{author}{\bibfnamefont{M.}~\bibnamefont{Mary{\v s}ko}},
  \bibinfo{author}{\bibfnamefont{S.}~\bibnamefont{Mat{\v e}jkov{\'a}}},
  \bibinfo{author}{\bibfnamefont{Z.}~\bibnamefont{Sofer}}, \bibnamefont{and}
  \bibinfo{author}{\bibfnamefont{M.}~\bibnamefont{Pumera}},
  \bibinfo{journal}{ACS Nano} \textbf{\bibinfo{volume}{7}},
  \bibinfo{pages}{5930} (\bibinfo{year}{2013}).

\bibitem[{\citenamefont{Overhauser}(1953)}]{PhysRev.92.411}
\bibinfo{author}{\bibfnamefont{A.~W.} \bibnamefont{Overhauser}},
  \bibinfo{journal}{Phys. Rev.} \textbf{\bibinfo{volume}{92}},
  \bibinfo{pages}{411} (\bibinfo{year}{1953}).

\bibitem[{\citenamefont{{Anderson} and {Freeman}}(1962)}]{1962JChPh..37...85A}
\bibinfo{author}{\bibfnamefont{W.~A.} \bibnamefont{{Anderson}}}
  \bibnamefont{and}
  \bibinfo{author}{\bibfnamefont{R.}~\bibnamefont{{Freeman}}},
  \bibinfo{journal}{\jcp} \textbf{\bibinfo{volume}{37}}, \bibinfo{pages}{85}
  (\bibinfo{year}{1962}).

\bibitem[{\citenamefont{Duplock et~al.}(2004)\citenamefont{Duplock, Scheffler,
  and Lindan}}]{PhysRevLett.92.225502}
\bibinfo{author}{\bibfnamefont{E.~J.} \bibnamefont{Duplock}},
  \bibinfo{author}{\bibfnamefont{M.}~\bibnamefont{Scheffler}},
  \bibnamefont{and} \bibinfo{author}{\bibfnamefont{P.~J.~D.}
  \bibnamefont{Lindan}}, \bibinfo{journal}{Phys. Rev. Lett.}
  \textbf{\bibinfo{volume}{92}}, \bibinfo{pages}{225502}
  (\bibinfo{year}{2004}).

\bibitem[{\citenamefont{Boukhvalov et~al.}(2008)\citenamefont{Boukhvalov,
  Katsnelson, and Lichtenstein}}]{PhysRevB.77.035427}
\bibinfo{author}{\bibfnamefont{D.~W.} \bibnamefont{Boukhvalov}},
  \bibinfo{author}{\bibfnamefont{M.~I.} \bibnamefont{Katsnelson}},
  \bibnamefont{and} \bibinfo{author}{\bibfnamefont{A.~I.}
  \bibnamefont{Lichtenstein}}, \bibinfo{journal}{Phys. Rev. B}
  \textbf{\bibinfo{volume}{77}}, \bibinfo{pages}{035427}
  (\bibinfo{year}{2008}).

\bibitem[{\citenamefont{Balog et~al.}(2010)\citenamefont{Balog, Jorgensen,
  Nilsson, Andersen, Rienks, Bianchi, Fanetti, Laegsgaard, Baraldi, Lizzit
  et~al.}}]{Balog:2010fk}
\bibinfo{author}{\bibfnamefont{R.}~\bibnamefont{Balog}},
  \bibinfo{author}{\bibfnamefont{B.}~\bibnamefont{Jorgensen}},
  \bibinfo{author}{\bibfnamefont{L.}~\bibnamefont{Nilsson}},
  \bibinfo{author}{\bibfnamefont{M.}~\bibnamefont{Andersen}},
  \bibinfo{author}{\bibfnamefont{E.}~\bibnamefont{Rienks}},
  \bibinfo{author}{\bibfnamefont{M.}~\bibnamefont{Bianchi}},
  \bibinfo{author}{\bibfnamefont{M.}~\bibnamefont{Fanetti}},
  \bibinfo{author}{\bibfnamefont{E.}~\bibnamefont{Laegsgaard}},
  \bibinfo{author}{\bibfnamefont{A.}~\bibnamefont{Baraldi}},
  \bibinfo{author}{\bibfnamefont{S.}~\bibnamefont{Lizzit}},
  \bibnamefont{et~al.}, \bibinfo{journal}{Nat Mater}
  \textbf{\bibinfo{volume}{9}}, \bibinfo{pages}{315} (\bibinfo{year}{2010}).

\bibitem[{\citenamefont{Pujari et~al.}(2011)\citenamefont{Pujari, Gusarov,
  Brett, and Kovalenko}}]{PhysRevB.84.041402}
\bibinfo{author}{\bibfnamefont{B.~S.} \bibnamefont{Pujari}},
  \bibinfo{author}{\bibfnamefont{S.}~\bibnamefont{Gusarov}},
  \bibinfo{author}{\bibfnamefont{M.}~\bibnamefont{Brett}}, \bibnamefont{and}
  \bibinfo{author}{\bibfnamefont{A.}~\bibnamefont{Kovalenko}},
  \bibinfo{journal}{Phys. Rev. B} \textbf{\bibinfo{volume}{84}},
  \bibinfo{pages}{041402} (\bibinfo{year}{2011}).

\bibitem[{\citenamefont{Hampel et~al.}(1998)}]{Hampel:1997fc}
\bibinfo{author}{\bibfnamefont{W.}~\bibnamefont{Hampel}} \bibnamefont{et~al.}
  (\bibinfo{collaboration}{GALLEX Collaboration}),
  \bibinfo{journal}{Phys.Lett.} \textbf{\bibinfo{volume}{B420}},
  \bibinfo{pages}{114} (\bibinfo{year}{1998}).

\bibitem[{\citenamefont{Abdurashitov et~al.}(1999)}]{Abdurashitov:1998ne}
\bibinfo{author}{\bibfnamefont{J.}~\bibnamefont{Abdurashitov}}
  \bibnamefont{et~al.} (\bibinfo{collaboration}{SAGE Collaboration}),
  \bibinfo{journal}{Phys.Rev.} \textbf{\bibinfo{volume}{C59}},
  \bibinfo{pages}{2246} (\bibinfo{year}{1999}), \eprint{hep-ph/9803418}.

\bibitem[{\citenamefont{Bellini et~al.}(2013)}]{Borexino:2013xxa}
\bibinfo{author}{\bibfnamefont{G.}~\bibnamefont{Bellini}} \bibnamefont{et~al.}
  (\bibinfo{collaboration}{Borexino Collaboration}), \bibinfo{journal}{JHEP}
  \textbf{\bibinfo{volume}{1308}}, \bibinfo{pages}{038} (\bibinfo{year}{2013}),
  \eprint{1304.7721}.

\end{thebibliography}

\end{document}